\documentclass[12pt,preprint]{aastex}

\def\ro{{\it ROSAT\/}}
\def\asca{{\it ASCA\/}}

\slugcomment{To appear in \apj}

\shorttitle{A High-Energy Study of the Geminga Pulsar}
\shortauthors{M. Jackson et al.}

\begin{document}

\title{A High-Energy Study of the Geminga Pulsar}

\author{M. S. Jackson, J. P. Halpern, and E. V. Gotthelf}
\affil{Department of Astronomy, Columbia University, New York, NY, 10025-6601}

\author{J. R. Mattox} 
\affil{Department of Physics \& Astronomy, Francis Marion University, Florence, SC 29501-0547}

\begin{abstract}
We present the results of deep X-ray and $\gamma$-ray observations of
the Geminga pulsar obtained in the final years of the \asca\ and {\it
CGRO} missions, and an upper limit from {\it RXTE\/}.  A
phase-connected ephemeris from the $\gamma$-rays is derived that spans
the years 1973--2000, after allowing for a minor glitch in frequency
of $\Delta f/f = 6.2 \times 10^{-10}$ in late 1996.  \asca\
observations of the hard X-ray pulse profile in 1994 and 1999 confirm
this glitch.  An improved characterization of the hard X-ray pulse
profile and spectrum from the long \asca\ observation of 1999 confirms
that there is a non-thermal X-ray component that is distinct from the
$\gamma$-ray spectrum as measured by EGRET.  It can be parameterized
as a power-law of photon index $\Gamma = 1.72 \pm 0.10$ with a flux of
$2.62 \times 10^{-13}$ ergs cm$^{-2}$ s$^{-1}$ in the $0.7-5$ keV band
and pulsed fraction $0.54 \pm 0.05$, similar to, but more precise than
values measured previously.  An extrapolation of this spectrum into
the energy band observed by the {\it RXTE\/} PCA is consistent with
the non-detection of pulsed emission from Geminga with that
instrument.  These results are interpreted in the context of outer-gap
models, and motivations for future X-ray observations of Geminga are
given.

\end{abstract}


\section{Introduction}

Discovered in 1972 by the SAS-2 satellite \citep{fi75,th77}, Geminga
is the second brightest $\gamma$-ray source in the sky above 100 MeV
\citep{sw81}. It was known only as a $\gamma$-ray source until a
promising candidate was detected in X-rays by the Einstein Observatory
\citep{bi83}, and associated with an optical counterpart
\citep{bi87,ht88,bi88}.  Subsequently, Geminga was found to be a
rotation-powered pulsar with a period of 237~ms in X-rays by \ro\
\citep{hh92}, and in $\gamma$-rays by the Energetic Gamma Ray
Experiment Telescope (EGRET) on the {\it Compton Gamma-ray
Observatory} \citep{be92}. Prior to the discovery of the 237~ms spin
period of Geminga, claims had been made for various periods in the
range 59--60~s in $\gamma$-rays and in X-rays
\citep{th77,ma77,zm83,bi84,zy88,ka85}, but no such detections have
been made in high quality X-ray and $\gamma$-ray observations during the
past decade.

The optical spectrum of Geminga is predominantly non-thermal, with
possible ion cyclotron features \citep{mhs98,mcb98}.  \cite{sh98}
reported optical modulation from Geminga that resembles its
$\gamma$-ray light curve.  Geminga is unusual as a rotation-powered
pulsar because it is not a strong radio source. In 1997, three groups
\citep{mm97,kl97,sp97} claimed detection of pulsed radio emission at
102~MHz, but observations at other radio frequencies have so far been
negative \citep{rdi98,mcl99,bfb99,kl99}.  A phase-connected ephemeris
covering the first 27 years of $\gamma$-ray observations of Geminga
was presented and updated by Mattox, Halpern, \& Caraveo (1998,
2000).

In this paper we present the results of a long observation with the
{\it Advanced Satellite for Cosmology and Astrophysics} (\asca ),
which allows us to better constrain the hard X-ray spectrum of Geminga
and perform pulse-phase spectroscopy.  X-ray pulse times of
arrival are compared with the latest ephemeris from EGRET.  Additional
constraints on the hard X-ray emission are derived from an observation
by the {\it Rossi X-ray Timing Explorer} Proportional Counter Array
({\it RXTE\/} PCA). 

\section{Observations}

A log of the observations used in this paper is given in Table \ref{tbl-1}.
A six-day observation of Geminga was obtained by \asca\ in 1999 
October 5--11.  Observations by {\it RXTE\/} were made in 1996 April and May.
EGRET made many $\gamma$-ray observations of Geminga since 1991,
until {\it CGRO} was de-orbited in 2000.  We also make brief use of
a 1993 \ro\ observation of Geminga and a 1994 observation by \asca,
both of which were described in detail in \cite{hw97}.

The \asca\ satellite \citep{tih94} incorporates four co-aligned focal
plane detectors, two Gas Imaging Spectrometers (GIS) and two
Solid-State Imaging Spectrometers (SIS).  Each instrument is
positioned at the focus of a conical foil mirror.  The SIS instrument
is sensitive to X-rays with energies between 0.4 and 10 keV and the
GIS detects X-rays between 0.7 and 10 keV. Only the GIS has time
resolution sufficient to study the pulsations of Geminga.  The 1999
observation was designed with bit assignments such that the time
resolution was 0.49~ms in high bit rate telemetry mode, and 3.91~ms at
medium bit rate.  This required the sacrifice of two bits from the
pulse-height analyzer resulting in 256 energy channels instead of the
default 1024, and elimination of the five bits of rise-time
information; both of these have negligible effect on the quality of
the data.  The GIS high and medium bit rate data were combined for the
timing and spectral analysis.  The small amount of low bit rate data
in the GIS were not used.  For the spectral analysis, data from both
the GIS and SIS instruments were used.  The SIS observations were
conducted in 1-CCD mode.

{\it RXTE\/} \citep{brs93} consists of the Proportional Counter Array
(PCA), High Energy X-ray Timing Experiment (HEXTE), and the All Sky
Monitor (ASM). The PCA consists of five xenon filled proportional
counters, coaligned and collimated, that observe a roughly circular
field with $\approx 1^\circ$ FWHM response \citep{jah96}. The RXTE
data analyzed here are exclusively from the PCA, because of its large
collecting area ($\sim6500$ cm$^2$ at 10 keV). The PCA data are of
interest to search for pulsed emission at higher X-ray energies than
\asca, since the PCA is sensitive to X-rays with energies between 2 and 60
keV.

EGRET is a high-energy $\gamma$-ray telescope which is sensitive in
the energy range from about 30 MeV to 30 GeV. It measures the energy
of incoming $\gamma$-rays with a NaI(Tl) crystal, and their direction
with a multilevel thin-plate spark chamber system which converts the
gamma-rays and determines the trajectories of the resulting $e^{\pm}$
pair, and a triggering telescope that detects the presence of a pair
having the correct direction of the motion.  An anti-coincidence
scintillation dome discriminates against charged particles.  EGRET's
effective area is $\approx 1000$~cm$^{2}$ at 150~MeV, 1500~cm$^{2}$
around $0.5-1$~GeV, decreasing gradually at higher energies.  We used
all of the $>100$ MeV photons detected by EGRET from Geminga between
1991 and 1997 to make a mean light curve. The event selection is
described in \cite{ma98}.

\section{Timing Analysis}

\subsection{The Drifting EGRET Ephemeris}

Gamma-ray timing provides the most precise and continuous rotational
ephemeris of Geminga because of the sharpness of its $\gamma$-ray
peaks and the large number of observations that have been performed
over the years.  Combined analysis of EGRET data up to 1997, and
earlier data from the {\it COS B\/} satellite, yielded a
phase-connected, cubic ephemeris \citep{ma98} that fitted the
rotational phase of Geminga for 24 years with residuals smaller than
0.05 cycles (the ``1997 ephemeris'').  We reproduce this ephemeris in
Table \ref{tbl-2}.  The residuals from this timing solution were
fitted by \cite{ma98} to a sinusoidal term of semiamplitude 0.026
cycles, possibly attributable to an orbiting planet, but it is now
apparent that this was simply a manifestation of timing noise.
Beginning in 1997, the phase measured by EGRET began to deviate
systematically from the pre-1997 ephemeris in the sense that the peaks
arrive earlier than expected \citep{ma00}.  

\begin{figure}
\center{\scalebox{0.65}{\plotone{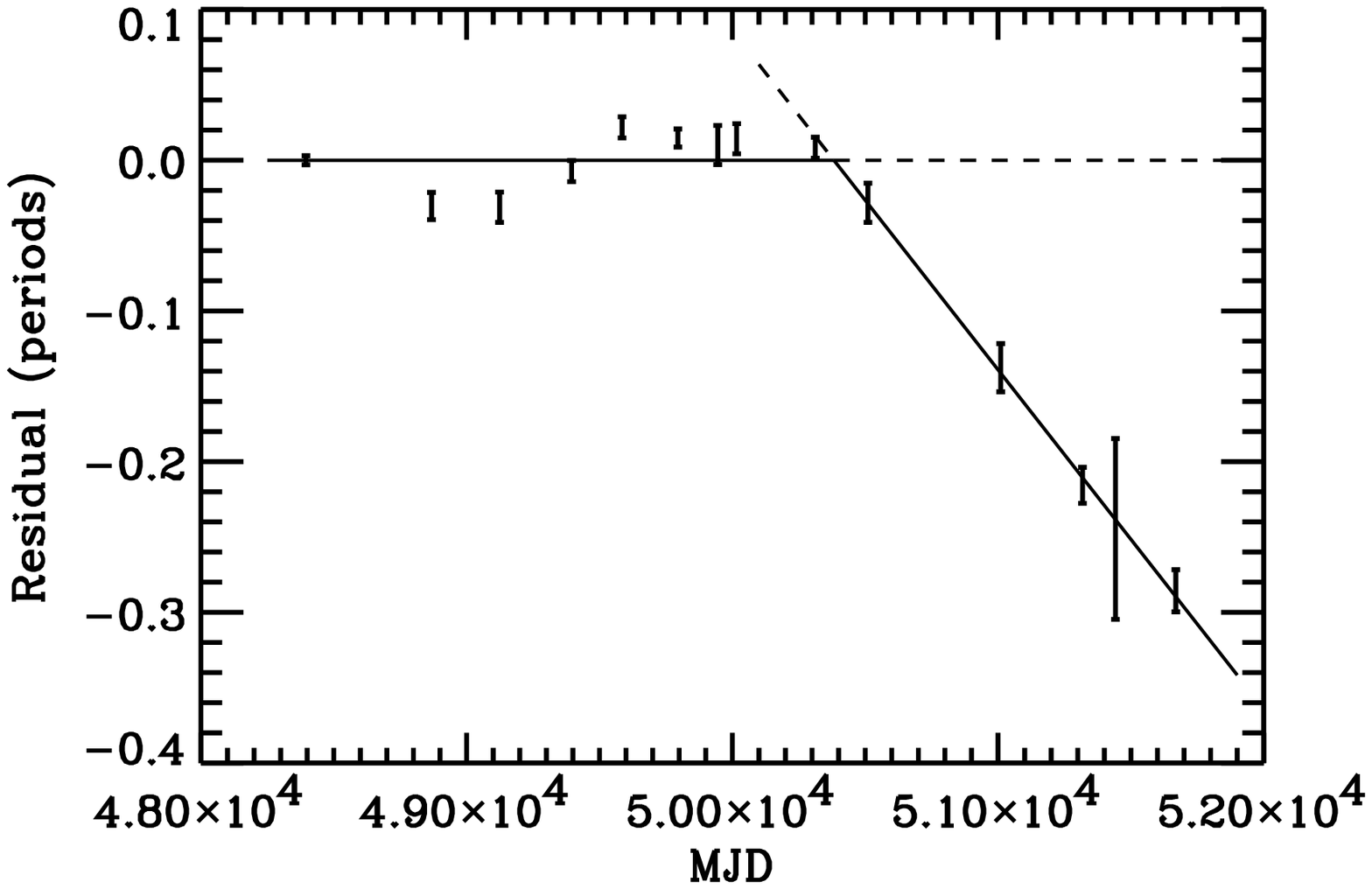}}}
\caption{Phase residuals of the EGRET timing observations of Geminga
relative to the cubic ``1997 ephemeris'' of \cite{ma98} (pre-glitch
ephemeris in Table \ref{tbl-2}).  The 14 measurements
correspond to the numbered EGRET
epochs in Table \ref{tbl-1}, some of which were grouped together.
The {\it solid line} represents the cubic ephemeris segments
before and after the glitch.
The post-glitch ephemeris is also given in Table
\ref{tbl-2}.\label{phaseplot}}
\end{figure}

Figure \ref{phaseplot} shows the phase residuals of all of the EGRET
observations relative to the 1997 ephemeris.  The points in Figure
\ref{phaseplot} correspond to the numbered EGRET epochs in Table
\ref{tbl-1}, some of which were grouped together.  According to a fit
of the light curves to Lorentzians, described in \cite{ma98}, in 1998
July, the gamma-ray peaks were observed to arrive $0.14 \pm 0.02$ of a
cycle earlier than predicted; in 1999 May they arrived $0.22 \pm 0.01$
cycles early; in 1999 September they arrived $0.24 \pm 0.06$ cycles
early; and during the final EGRET observation in 2000 April--May, the
gamma-ray peaks arrived $0.29 \pm 0.01$ cycles early.  This departure
from the phase predicted by the 1997 ephemeris is consistent with a
minor glitch (discontinuous change in frequency) shortly after the
time of the 1996 measurement, MJD $\approx 50382$ (1996 Oct. 26), by
an amount $\Delta f/f = 6.2 \times 10^{-10}$.  By fitting the phase
residuals in Figure \ref{phaseplot}, we derive an approximate
post-glitch ephemeris, also given in Table \ref{tbl-2}.  Hereafter, we
refer to the ``EGRET ephemeris'' as the two cubic segments covering
the years 1991--2000 in Table \ref{tbl-2}, plus the residuals from the
cubic terms that are illustrated in Figure \ref{phaseplot}.  In the
following analysis, we interpolate the EGRET ephemeris, including its
phase residuals, in order to phase-align and compare pulse profiles
from X-ray wavelengths.

\subsection{ASCA Observation}

All of the event times in the \asca\ GIS observation of 1999 October
were first converted to Barycentric Dynamical Time (TDB) using the
position and proper motion of Geminga listed in Table \ref{tbl-2}.
This was done using the ``timeconv'' program, which is part of the
FTOOLS software package.

\begin{figure}
\scalebox{0.85}{\rotatebox{270}{\plotone{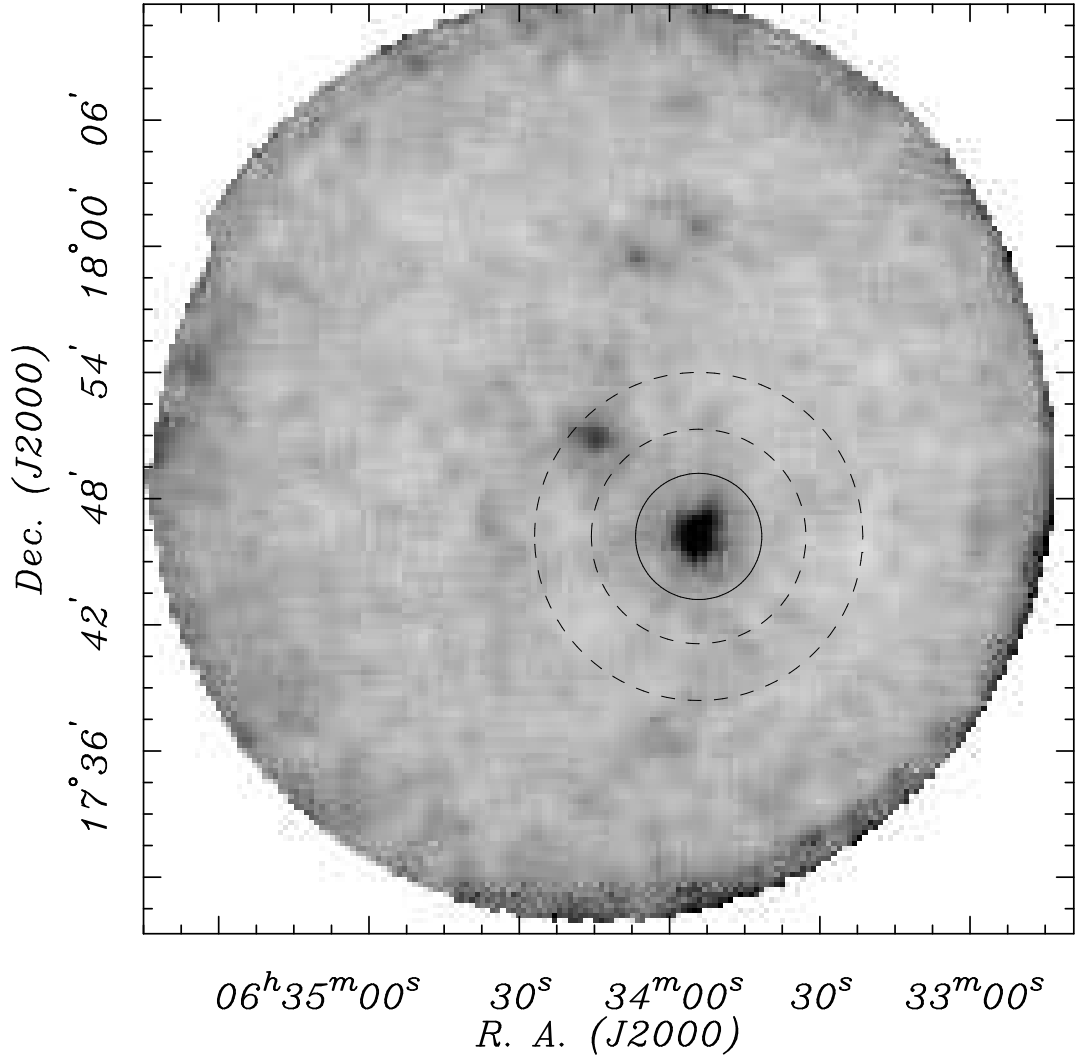}}}
\caption{\asca\ GIS image of Geminga, showing the source circle (solid line) and annulus (dashed lines)
from which the background was subtracted for calculation of the pulse
profile. The radii of the source circle
($3^{\prime}$) and background annulus ($5^{\prime}-6.\!^{\prime}25$) were
chosen to maximize the signal-to-noise ratio in the light curve.
\label{point1}}
\end{figure}

We determined optimal radii for the source extraction and background
annulus in the GIS by maximizing the signal-to-noise ratio in the
resulting light curve.  As shown in Figure \ref{point1}, the radius of the source
circle was chosen to be $3^{\prime}$, and a concentric annulus of inner
and outer radii of $5^{\prime}$ and $6.\!^{\prime}25$, respectively
gave a good estimate of the background.  We confirm that, as found by
\cite{be99} using previous \ro\ and \asca\ images, there is no evidence
for diffuse emission (synchrotron nebulosity) associated with Geminga.

\begin{figure}
\center{\scalebox{0.65}{\rotatebox{270}{\plotone{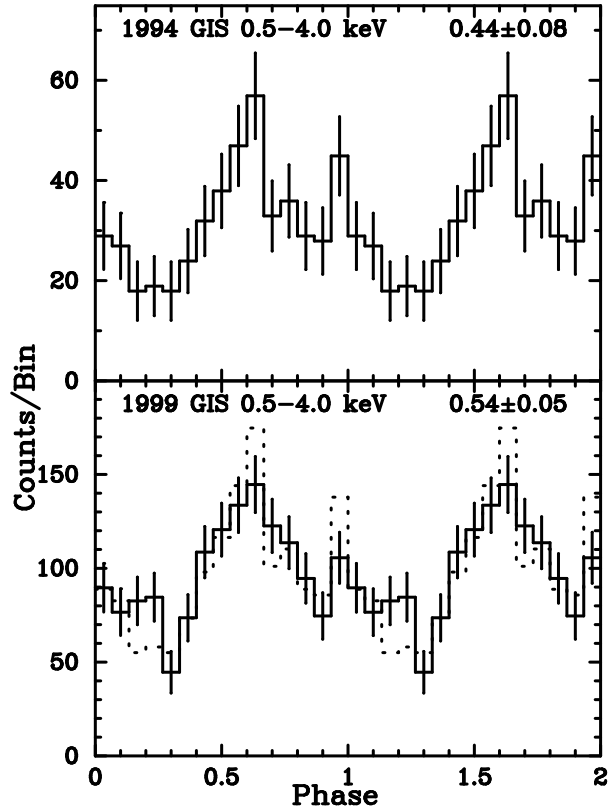}}}}
\caption{Comparison of background-subtracted
pulse profiles in the $0.5-4.0$~keV band from 1994 and 1999 \asca\ GIS
observations of Geminga.
The dashed line is the 1994 light curve normalized to the total counts
in the 1999 light curve.
Phase zero corresponds to epoch $T_0$ in Table \ref{tbl-2}.
The pulsed fractions and their uncertainties are indicated in
each panel. \label{pulse1}}
\end{figure}

In Figure \ref{pulse1} we compare the resulting $0.5-4.0$ keV light
curve with the only previous hard X-ray light curve of Geminga made
with the same instrument in 1994 March and described by \cite{hw97}.
The curves have been aligned according to the EGRET ephemeris, and the
resulting agreement in phase confirms the drifting EGRET ephemeris
shown in Figure \ref{phaseplot}.  To evaluate whether the pulse shape
experienced any change between the 1994 and 1999 \asca\ observations,
the two light curves in Figure \ref{pulse1} were compared using
$\chi^2$ test, after accounting for the difference in the exposure
times.  It was determined that the light curves do not differ
significantly. The stability of the light curve, and its large pulsed
fraction and strong main peak allow the possibility of continuing the
rotational ephemeris of Geminga using hard X-ray observations, e.g.,
with {\it XMM-Newton} and {\it Chandra}, during the current epoch in
which there are no high-energy $\gamma$-ray instruments in orbit.

\begin{figure}
\center{\scalebox{0.5}{\plotone{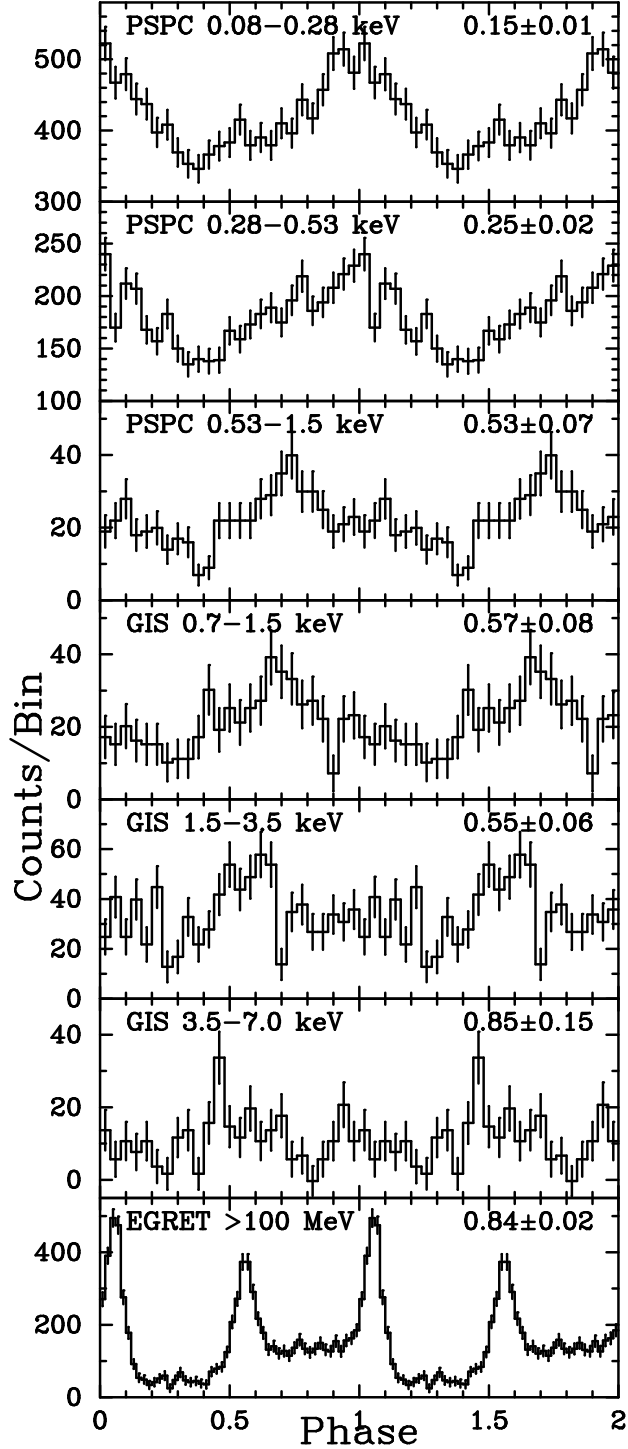}}}
\caption{Pulse profiles from 1993 \ro\ PSPC and 1999 \asca\ GIS
observations of Geminga, along with the summed EGRET light curve,
folded at the ephemeris of Table \ref{tbl-2}.
Phase zero corresponds to epoch $T_0$ in Table \ref{tbl-2}.
The pulsed fractions and their uncertainties are indicated in
each panel.  Background subtraction of all 
light curves has been performed. Compare
with the similar Figure 9 of \cite{hw97}.\label{pulse2}}
\end{figure}

Figure \ref{pulse2} shows the folded light curves for the 1999 GIS
data divided into three energy bands, $0.7-1.5$~keV, $1.5-3.5$~keV,
and $3.5-7.0$~keV.  For comparison, we also reproduce soft X-ray pulse
profiles from a 1993 September observation with the \ro\ PSPC that was
published by \cite{hw97}, in energy bands $0.08-0.28$~keV,
$0.28-0.53$~keV, and $0.53-1.50$~keV.  The summed EGRET light curve
above 100~MeV is also shown. For CGRO viewing period 1, comparing the
number of events selected with E$>$100 MeV to the likelihood estimate
of Geminga flux \citep{ma96} 63\% of the events selected are estimated
to be from Geminga.  The remaining 37\% would be primarily diffuse
Galactic gamma-ray emission. This is the estimated background for
the EGRET lightcurve. It is statistically consistent with the claim in
\cite{mh94} that there is negligible unpulsed emission from Geminga in
that energy band. These light curves resemble closely those given in
Figure 9 of \cite{hw97}. Because the 1999 GIS observation had 2.5
times more exposure time than the 1994 observation, the GIS light
curves could be split into smaller energy bands, yielding more
information about the change in pulse profile from low to medium X-ray
energies than was available before.  It was clear from the analysis of
\cite{hw97} that thermal surface emission, fitted with $T_{\rm BB} =
(5.6 \pm 0.5) \times 10^5$~K, dominates the X-ray emission below 0.7
keV, whereas the \asca\ GIS, which is sensitive above 0.7 keV, sees
primarily non-thermal emission (see also \S 4).  The large pulsed
fractions of the light curves above 0.7~keV bear out this
interpretation.  In addition, it was noted in \cite{hw97} that the
non-thermal pulse component itself has energy dependent structure, in
particular, an offset of 0.1 cycles between the $0.7-1.5$~keV PSPC and
$1.0-4.0$~keV GIS light curves.  Figures \ref{pulse1} and \ref{pulse2}
confirm that the GIS light curve leads the hardest PSPC band
($0.53-1.50$~keV) by approximately 0.1 cycles. However, since the
energy resolution of the PSPC is so poor, it is possible that some
soft photons of thermal origin bias this light curve's peak toward
later times.

The non-thermal X-ray pulse as illustrated in Figures \ref{pulse1} and
\ref{pulse2} is best described as a single broad peak centered at
phase 0.63, approximately coinciding with the smaller of the two EGRET
peaks, with a possible minor peak at phase 0.97.  At the highest
energy ($>3.5$~keV), the GIS light curve may begin to resemble the
EGRET peaks in having two peaks separated by 0.5 cycles, but the
statistics are poor and, even if real, the GIS peaks lead the EGRET
peaks by $\sim 0.1$ cycles.  On balance, the differences between the
GIS and EGRET light curves seem greater than their similarities, and
we regard this as evidence that the hard X-rays are not simply an
extension of the $\gamma$-ray spectrum (see also \S 4).

\subsection{RXTE Observation}

The PCA instrument on RXTE was used to search for pulsed emission from
$2-60$~keV. The data were recorded in Good Xenon mode and processed
using standard methods. The time resolution for the observation of
$0.95 \mu$s and absolute clock precision of $5 \mu$s are more than
adequate to distinguish pulsations from Geminga. The energy
information was stored in eight bits in the event word, and the
energy-channel conversion table for Cycle 3 was used to convert to
energy in keV.

The barycenter-corrected PCA data in the energy ranges $2-15$~keV and
$15-60$~keV were folded using the EGRET ephemeris.  The faseBin
routine, an epoch folding test, a Fourier transform pulse search, a
Rayleigh test \citep{le83}, and the $Z^2_n$ test of orders 2 and 3
\citep{bu83} were performed on the data.  No significant pulsations
were detected using any of these techniques in the energy range in
which pulsations were detected by \asca, nor at higher energies, with
a flux upper limit of $3.5 \times 10^{-13}$ ergs cm$^{-2}$ s$^{-1}$ in
the $2-15$ keV energy range . This negative result is undoubtedly due
to the high background that dominates in this non-imaging instrument,
in combination with spectral shape of Geminga's non-thermal emission,
as will be discussed in the next section.

\section{Spectral Analysis}

\subsection{ASCA X-ray Spectrum}

The FTOOLS/XSELECT software package was used to prepare the \asca\
data for spectral analysis. Source spectra were extracted from a
circle $2.\!^{\prime}5$ in radius for the SIS data, and
$3.\!^{\prime}5$ in radius for the GIS data.  The source was at a
suitable location for background spectra to be extracted from annuli
concentric with the source circle for both SIS and GIS. It is
important to choose an annulus close enough to the source circle to
acquire a fair sample of the background sky at the point of the
source, but not close enough to contain a significant number of source
counts, which would oversubtract the source spectrum.

To obtain response matrices, the latest GIS ``rmf'' files were
downloaded and rebinned to match the reduced number of energy channels
in the data. The ``sisrmg'' program was used to generate response
matrices (rmfs) for the SIS spectral files.  Ancillary response (arf)
files were created from the rmf files using the program ``ascaarf''.
The spectral data were rebinned, using the program ``grppha'' so that
there were 40 counts or more in each bin.  This reduces the noise,
especially at higher energies where the instrument has less
sensitivity, and allows Gaussian statistics to be used in the
evaluation of the fit.

\begin{figure}
\scalebox{0.65}{\rotatebox{270}{\plotone{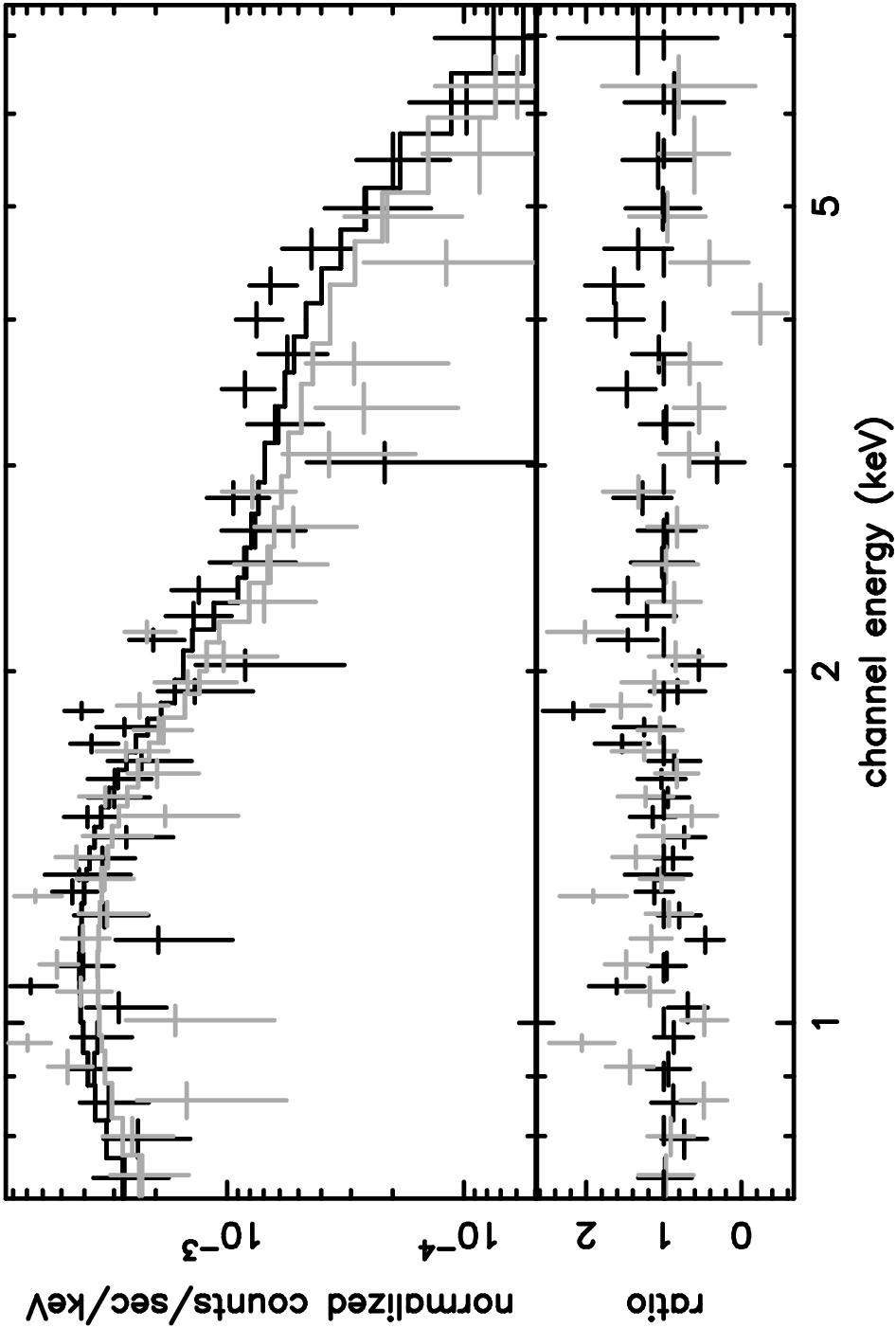}}}
\caption{Fit of \asca\ SIS spectra of Geminga to a power law.
{\it Dark line}: The SIS0 data and model;
{\it Light line}: The SIS1 data and model.
The lower panel shows the residuals of the data from the model.
The fitting parameters are given in Table \ref{tbl-3}. \label{sisspect}}
\end{figure}

\begin{figure}
\scalebox{0.65}{\rotatebox{270}{\plotone{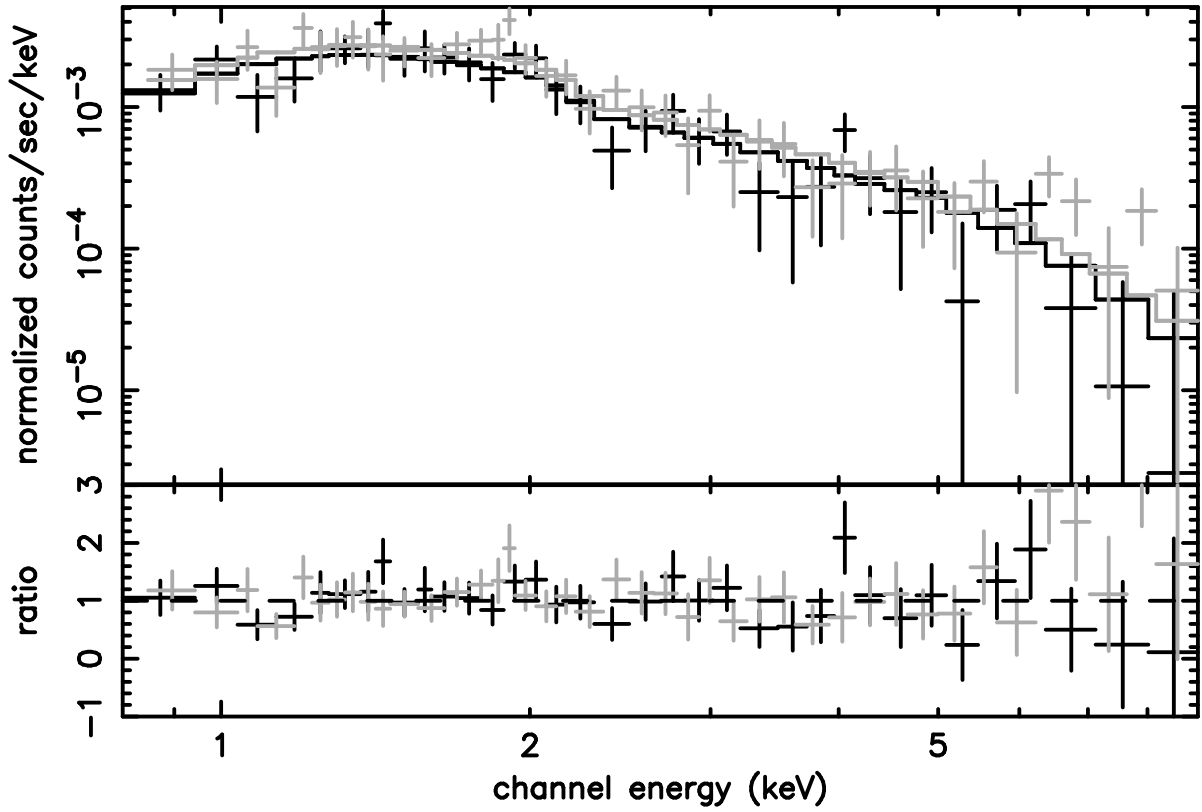}}}
\caption{Fit of \asca\ GIS spectra of Geminga to a power law.
{\it Dark line}: The GIS2 data and model;
{\it Light line}: The GIS3 data and model.
The lower panel shows the residuals of the data from the model.
The fitting parameters are given in Table \ref{tbl-3}. \label{gisspect}}
\end{figure}

Guided by the combined \ro/\asca\ results of \cite{hw97}, we fitted
simple power-law models to each of the \asca\ instruments using XSPEC.
The soft blackbody component that dominates the \ro\ spectrum does not
contribute significantly in the \asca\ band, and was neglected.  In
this analysis, the intervening column density $N_{\rm H}$ was held
fixed at the best fitted value of $1.07\times 10^{20}$ given in
\cite{hw97}.  Uncertainties in this small $N_{\rm H}$ do not affect
the \asca\ results.  The energy limits of the fit for the SIS data
were set at $0.7-10$ keV for SIS0, and $0.7-7.5$ keV for SIS1 data
because of its poorer signal-to-noise at higher energy. The GIS
spectra were fitted from $0.8-10$ keV to avoid calibration
uncertainties at the lowest energies.  In addition to fits of the
individual instruments, combined fits were made.  The combined SIS and
GIS fits are shown in Figures \ref{sisspect} and \ref{gisspect}.

Results of the fits are given in Table \ref{tbl-3} for the combined
SIS and GIS as well as for all instruments combined. All fits are
acceptable, with $\chi^2\sim 1$ per degree of freedom.  Although not
all of the instruments are consistent with each other at the $68\%$
confidence level, three out of four are consistent with the combined
SIS+GIS fit, which yields photon index $\Gamma =
1.72^{+0.10}_{-0.09}$.  This compares favorably with the lower-quality
results of \cite{hw97}, who found $\Gamma = 1.47^{+0.25}_{-0.23}$ for
SIS0 and $\Gamma = 2.19^{+0.35}_{-0.31}$ for GIS3 in the 1994
observation.

We also performed crude pulse-phase spectroscopy by separating the
GIS events into two phase bins, 0.28--0.88 in Figures \ref{pulse1}
and \ref{pulse2} (which includes most of the main peak),
and the remainder (0.88--0.28).  Power-law fits in the range
$0.8-8.0$~keV were performed for both sets, and the results are
also listed in Table \ref{tbl-3}.  While the spectrum of the
main peak is slightly softer than that of the interpeak region
($\Gamma = 2.00 \pm 0.24$ vs. $\Gamma = 1.74 \pm 0.24$), the
difference is only marginally significant.

Table \ref{tbl-3} also gives the modeled fluxes in the range $0.7-5.0$
keV, normalized to a full cycle. The fluxes for the full data sets are
similar to those for the PSPC+SIS0 fit given in \cite{hw97}, but are
more than twice the flux in the PSPC+GIS3 fit given there. The fluxes
for the individual instruments in our 1999 \asca\ observation agree
with each other quite well, with the GIS tending to show a slightly
higher flux than the SIS.  We conclude that the discrepancies between
SIS and GIS spectra in the 1994 observation noted by \cite{hw97} were
largely due to the shorter exposure and non-optimal mode of that
observation, while the current results from the 1999 observation are
more internally consistent and precise.

\subsection{Broad-band Spectrum of Geminga}

\begin{figure}
\center{\scalebox{1.}{\plotone{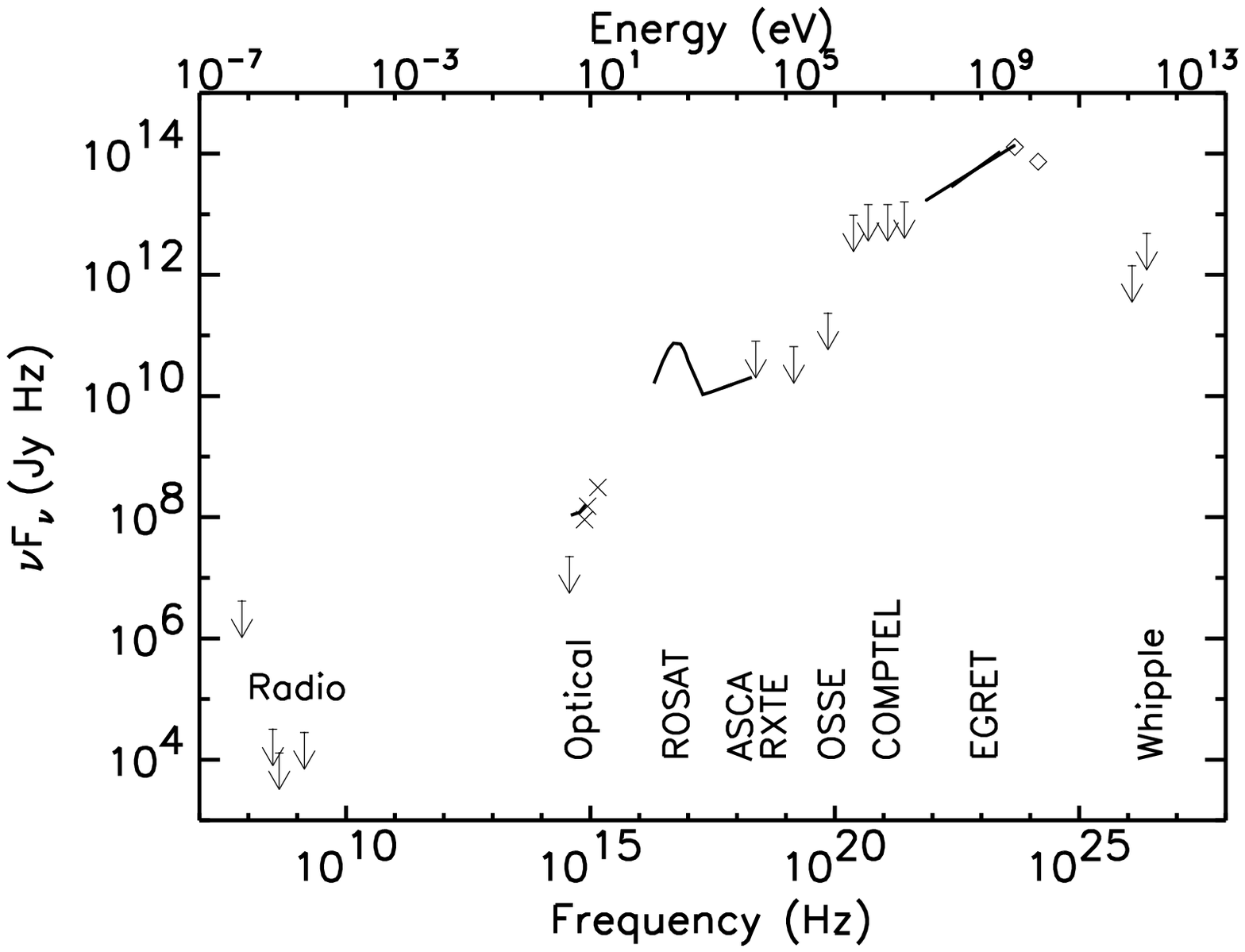}}}
\caption{The broad-band energy spectrum of Geminga. Upper limits
correspond to that of the pulsed flux, whereas the data points
represent the total flux. In order of increasing energy, the
references are as follows: The upper limits in the radio are found in
\cite{bfb99} and \cite{kl99}. The $I$-band upper limit is from
\cite{bi96}. The optical spectrum is from \cite{mhs98}. The UV points
({\it X's}) are from \cite{mcb98} and \cite{bi96}. The curve in the
soft X-ray band represents the blackbody plus power-law fit to the
\ro\ PSPC \citep{hw97}. The \asca\ curve represents the power law
fitted to the \asca\ data in this paper. The upper limits at 10 and 60
keV (just above the \asca\ points) represent the pulsed detection
threshold at those energies of the {\it RXTE\/} data described in the
text, assuming a continuation of the \asca\ measured power-law index.
The upper limit at 300 keV is from OSSE
\citep{sch95} and the COMPTEL upper limits at 1$-$11 MeV are from
\cite{ku96}. The EGRET $\gamma$-ray curves are from \cite{mh94} and
\cite{fi98}, respectively, and the points just above the curves are
also from \cite{fi98}. The VHE upper limits are from
\cite{ak93}. \label{fullspect}}
\end{figure}

An updated broad-band spectrum of Geminga, including the new data
presented in this paper, is shown in Figure \ref{fullspect}.  The
previous sources for data and upper limits are given in the Figure
caption.  We derived upper limits from {\it RXTE\/} for the pulsed
flux by determining the minimum pulsed detection threshold for the
$Z^2_2$ test (which should provide the greatest sensitivity of all
$Z^2_n$ tests, for a pulse shape similar to the hard X-ray or
$\gamma$-ray pulse profile in Figure \ref{pulse2}) and the epoch
folding test. The frequency at the epoch of the observation is
well-defined by the EGRET ephemeris (i.e., a search over frequency is
not required, although we nevertheless performed a search in a small
frequency range immediately surrounding the expected value), and the
pulse shape can be approximated by a sine wave.  Using the appropriate
parameters similar to those in \cite{le83} for the epoch folding test
(which, for a pulse shape like that at the highest energy observed by
\asca\, has the lowest pulsed detection threshold), we derived upper
limits for 10 and 60 keV corresponding to the collecting area of the
PCA under the assumption of a power-law of $\Gamma = 1.72$ as measured
by \asca. It is apparent from Figure \ref{fullspect} that, given its
flux and spectral index, Geminga is too faint for its pulsed flux to
have been detected at any energy by the {\it RXTE\/} PCA instrument.

\section{Interpretation}

It is clear from Figure \ref{fullspect} that the $\gamma$-ray emission
from Geminga represents the energetically dominant component of the
total emission of the pulsar, while an extrapolation of the nonthermal
X-ray spectrum falls well below the EGRET flux.  The hard X-rays
therefore represent a physically distinct emission mechanism from the
primary $\gamma$-rays.  A general theory of X-ray emission from
rotation-powered pulsars \citep{wa98} predicted that any strong
$\gamma$-ray pulsar will also have a hard X-ray spectrum with photon
index $\Gamma = 1.5$.  In this model, an outer-gap accelerator will
send $e^{\pm}$ pairs flowing inward and outward along open magnetic
field lines.  These particles continuously radiate $\gamma$-rays by
the curvature mechanism.  When the inward flowing particles approach
the surface of the star, the $>100$~MeV $\gamma$-rays that they emit
convert into secondary $e^{\pm}$ pairs in the inner magnetosphere
wherever $B\,{\rm sin}\phi > 2 \times 10^{10}$~G, where $\phi$ is the
angle between the photon and the {\bf B} field.  Those secondary pairs
must radiate away their energy instantaneously in the strong local
{\bf B} field. Such a synchrotron decay spectrum has $\Gamma = 1.5$
between $\sim$0.2~keV and 5~MeV.  The luminosity of such a component
in the \asca\ bandpass is estimated theoretically from the
Goldreich-Julian current in the outer-gap accelerator, and is found to
be $\sim 1 \times 10^{30}$ ergs~s$^{-1}$, in agreement with the
observed flux from Geminga.

A three-dimensional magnetosphere model of Geminga with a thick outer
gap is presented by \cite{zc01}, who explain the phasing of the soft
and hard X-ray emission with respect to the the $\gamma$-ray pulse
profile as observed by \cite{hw97}.  Their modeled hard X-ray peak
coincides with the first $\gamma$-ray peak as a result of the inward
motion of the X-ray synchrotron emitting particles, while the soft,
thermal X-ray peak coming from near the polar cap coincides with the
second $\gamma$-ray peak.  \cite{zc01} fitted the $\gamma$-ray pulse
profiles with a magnetic inclination angle of $\sim 50^{\circ}$ and a
viewing angle of $\sim 86^{\circ}$, which would also be compatible
with the pulse modulation of the soft X-rays.

While the fitted \asca\ spectral index of $\Gamma = 1.72 \pm 0.10$
does not strongly support or refute this theory, it can be noted that
a power-law fit is merely a parameterization of the shape of the X-ray
spectrum that does not allow for possible complications due to
additional processes that may be occurring.  For example, the
cyclotron energy for electrons in the pulsar's inner magnetosphere is
in the \asca\ bandpass, which means that the cyclo-synchrotron
spectrum must deviate from a power law there.  The nonthermal spectrum
must turn down at low energies, both from theoretical arguments, and
empirically in order not to exceed the observed ultraviolet and
visible flux.  Furthermore, it is possible that there is an
additional, hotter thermal component around 1~keV which would arise
from the polar caps heated by the impacting charged particles flowing
in along the open field lines, such as in the model of \cite{phh01}.
The latter model was based on double-blackbody fits to the \ro\
spectrum alone, taken from \cite{hr93}, but since superseded by
evidence from \asca\ that the harder component is non-thermal
\citep{hw97}.  Nevertheless, it is still possible that a fainter, hot
thermal component is present that is not easily distinguished from the
dominant non-thermal spectrum and pulse profiles, but that
contaminates the power-law fit.  Our pulse-phase spectroscopy, even in
the long \asca\ observation of 1999, is not sensitive enough to
meaningfully constrain such a complex model.  The detection of X-rays
above 10~keV from Geminga is an unmet challenge, but one that might be
necessary to better measure the the predicted power-law spectrum.

\section{Final Remarks}

We were startled to see a recent report of detection of a 62~s period
from Geminga in TeV $\gamma$-rays \citep{ns01}.  That paper cited nine
other detections of periods in the range 59--60~s (see \S 1 for
references), going all the way back to analyses of {\it SAS 2}, {\it
COS B}, {\it Einstein}, and {\it EXOSAT} data, none of which was ever
considered convincing \citep{bu85}.  We have searched \ro\ and \asca\
data in this range of periods without detecting significant
modulation.  For example, there is no signal at a period of 62.4~s
that would fit the linear extrapolation of previous claims to the
\asca\ observation of 1999.  We regard the case of the moving 60~s
period as closed a decade ago, notwithstanding this recent attempt at
revival.

Although there is no $\gamma$-ray telescope currently in orbit, it
will be possible to maintain a phase-connected ephemeris and monitor
Geminga's glitch activity using X-ray instruments such as {\it
Chandra} and {\it XMM-Newton}, now that we have confirmed the precise
phase relationship between the hard X-ray and $\gamma$-ray pulse in
two \asca\ observations. The standard dipole braking index $n = \ddot
f f/\dot f^2 = 3$ would correspond to $\ddot f$ smaller than the
observed value by a factor of 5.5.  Despite the existence of a 27~yr
long phase-connected $\gamma$-ray ephemeris of Geminga, it has not
been possible to measure a true braking index because $\ddot f$ is
apparently dominated by timing noise, even apart from the observed
glitch, which is among the smallest ever detected in a pulsar.  More
sensitive hard X-ray observations, such as are possible with {\it
XMM-Newton}, also have the potential to disentangle any additional
spectral and pulse components that may be present, such as cyclotron
features, hot polar cap emission, and even atomic spectral lines from
the neutron star photosphere.

\acknowledgments

\clearpage

\clearpage

\clearpage 

\clearpage 

\clearpage 

\clearpage 

\clearpage 

\clearpage 

\begin{deluxetable}{llrr}
\tabletypesize{\scriptsize}
\tablecaption{Log of Observations \label{tbl-1}}
\tablewidth{0pt}
\tablehead{
\colhead{Instrument} & \colhead{Dates} & \colhead{Exposure time (ks)} &
\colhead{Count rate (s$^{-1}$)} }
\startdata
EGRET (1) & 1991 Apr 22--May 7 & 1209.6 & 1.8$\times 10^{-3}$ \\
EGRET (1) & 1991 May 16--30 & 1209.6 & 1.6$\times 10^{-3}$ \\
EGRET (1) & 1991 Jun 8--15 & 604.8 & 1.5$\times 10^{-3}$ \\
EGRET (2) & 1992 Jun 11--25 & 1209.6 & 2.2$\times 10^{-4}$ \\
EGRET (2) & 1992 Aug 11--20 & 777.6 & 1.9$\times 10^{-4}$ \\
EGRET (2) & 1992 Sep 1--17 & 1382.4 & 1.7$\times 10^{-4}$ \\
EGRET (2) & 1992 Oct 8--15 & 604.8 & 1.4$\times 10^{-4}$ \\
EGRET (2) & 1992 Nov 3--17 & 1209.6 & 1.0$\times 10^{-4}$ \\
EGRET (3) & 1993 Mar 23--29 & 604.8 & 3.2$\times 10^{-4}$ \\
EGRET (3) & 1993 May 13--24 & 950.4 & 4.1$\times 10^{-4}$ \\
EGRET (4) & 1993 Dec 1--13 & 1036.8 & 3.5$\times 10^{-4}$ \\
EGRET (4) & 1994 Feb 8--17 & 777.6 & 7.7$\times 10^{-4}$ \\
EGRET (5) & 1994 Aug 9--29 & 1814.4 & 3.7$\times 10^{-4}$ \\
EGRET (6) & 1995 Feb 28--Mar 21 & 1814.4 & 4.0$\times 10^{-4}$ \\
EGRET (6) & 1995 Apr 4--11 & 604.8 & 2.5$\times 10^{-4}$ \\
EGRET (6) & 1995 May 9--Jun 6 & 2419.2 & 1.8$\times 10^{-4}$ \\
EGRET (7) & 1995 Aug 8--22 & 1209.6 & 2.4$\times 10^{-4}$ \\
EGRET (8) & 1995 Oct 17--31 & 1209.6 & 2.7$\times 10^{-4}$ \\
EGRET (9) & 1996 Jul 30--Aug 27 & 2419.2 & 3.4$\times 10^{-4}$ \\
EGRET (10) & 1997 Feb 18--Mar 18 & 2419.2 & 1.5$\times 10^{-4}$ \\
EGRET (11) & 1998 Jul 7--21 & 1209.6 & 2.0$\times 10^{-4}$ \\
EGRET (12) & 1999 May 11--25 & 1209.6 & 2.1$\times 10^{-4}$ \\
EGRET (13) & 1999 Sep 14--28 & 1209.6 & 4.7$\times 10^{-5}$ \\
EGRET (14) & 2000 Apr 25--May 9 & 1209.6 & 2.6$\times 10^{-4}$ \\
ROSAT PSPC & 1993 Sep 16--20 & 36.9 & 0.503 \\
ASCA SIS & 1994 Mar 28--31 & 49.2 & 1.7$\times 10^{-2}$ \\
ASCA GIS & 1994 Mar 28--31 & 75.3 & 1.2$\times 10^{-2}$ \\
ASCA SIS & 1999 Oct 5--11 & 194.0 & 1.4$\times 10^{-2}$ \\
ASCA GIS & 1999 Oct 5--11 & 207.8 & 1.0$\times 10^{-2}$ \\
RXTE PCA & 1996 Apr 27--May 8 & 142.7 & ---\\
\enddata

\end{deluxetable}
\clearpage 

\begin{deluxetable}{lll}
\tabletypesize{\scriptsize}
\tablecaption{The Geminga EGRET Ephemeris \label{tbl-2}}
\tablewidth{0pt}
\tablehead{\colhead{Parameter} & \colhead{Pre-glitch\tablenotemark{a}} & \colhead{Post-glitch\tablenotemark{b}}}
\startdata
Epoch of ephemeris $T_0$ (MJD\tablenotemark{c} )\tablenotemark{d}  & 46599.5 & 50381.999999364\\
Range of valid dates (MJD)  &  $41725-50382$ & $50382-51673$ \\
Frequency $f$ (Hz) & $4.217705363081(13)$ & $4.21764157512(18)$\\
Frequency derivative $\dot f$ (Hz s$^{-1}$) & $-1.9521712(12) \times 10^{-13}$ & $-1.951684(77) \times 10^{-13}$ \\
Frequency second derivative, $\ddot f$ (Hz s$^{-2}$) & $1.49(3) \times 10^{-25}$ & $1.49(3) \times 10^{-25}$ \\
   \tableline
   \multicolumn{1}{c}{Parameter\tablenotemark{e}} &\multicolumn{1}{c}{Value} & \\
   \tableline
Epoch of position (MJD)	& 49793.5 	& \\
R.A. (J2000) &	$6^{\rm h}33^{\rm m}54.\!^{\rm s}153$ & \\
Decl. (J2000) &	$+17^{\circ}46^{\prime}12.\!^{\prime\prime}91$ &  \\
R.A. proper motion $\mu_{\alpha}$ (mas yr$^{-1}$)	&  138 & \\
Decl. proper motion $\mu_{\delta}$ (mas yr$^{-1}$)	&  97  & \\
\enddata
\tablenotetext{a}{From \cite{ma98}.}
\tablenotetext{b}{$\dot f$ and $\ddot f$ assumed unchanged at the time of the glitch.}
\tablenotetext{c}{MJD=JD-2400000.5}
\tablenotetext{d}{Epoch of phase zero in Figures \ref{pulse1} and \ref{pulse2}.}
\tablenotetext{e}{Postion and proper motion from \cite{ca98}.}
\tablecomments{Digits in parentheses following a parameter value indicate $\sim$95\% confidence uncertainties 
in the last digits of the parameter.}
\end{deluxetable}

\clearpage

\begin{deluxetable}{lrr}
\tabletypesize{\scriptsize} 
\tablecaption{Power-law Fits to ASCA Spectra \label{tbl-3}}
\tablewidth{0pt} 
\tablehead{
\colhead{Instrument} & \colhead{$\Gamma$ (68\% conf. errors)} &
\colhead{Flux (ergs cm$^{-2}$ s$^{-1}$)\tablenotemark{a}}}
\startdata 
SIS &1.68 ($+$0.13, $-$0.12) & 2.29$\times 10^{-13}$\\
GIS &1.99 ($+$0.14, $-$0.14) & 2.78$\times 10^{-13}$\\
GIS+SIS &1.72 ($+$0.10, $-$0.09) & 2.62$\times 10^{-13}$ \\
GIS\tablenotemark{b} &2.00 ($+$0.24, $-$0.24)  & 3.30$\times 10^{-13}$\\
GIS\tablenotemark{c} &1.74 ($+$0.24, $-$0.23)  & 2.71$\times 10^{-13}$\\
\enddata 
\tablenotetext{a}{Flux in the $0.7-5.0$ keV range,
before interstellar absorption. In the case of phase-resolved
spectra, the flux has been normalized to a full cycle.}
\tablenotetext{b}{Phases 0.28--0.88 in Figures \ref{pulse1} and \ref{pulse2}.}
\tablenotetext{c}{Phases 0.88--0.28 in Figures \ref{pulse1} and \ref{pulse2}.}

\end{deluxetable}

\clearpage

\end{document}